\documentclass[nofootinbib,notitlepage,superscriptaddress,10pt,aps,pra,twocolumn]{revtex4-1}

\usepackage[utf8x]{inputenc}
\usepackage{amsmath}
\usepackage{amssymb}
\usepackage{graphicx}
\usepackage[normalem]{ulem}
\usepackage{epsf,epsfig}
\usepackage{psfrag}
\usepackage{color}
\usepackage{multirow}
\usepackage{diagbox}

\makeatletter
\newcommand\stroke[1]{\mathpalette\stroke@aux{#1}}
\def\stroke@aux#1#2{%
  \ooalign{%
    \hfil$#1^{\;\, \_\hspace{-0.05cm}\_}$\hfil\cr
    \hfil$#1#2$\hfil\cr
  }%
}
\makeatother
\newcommand\dtagliato{\stroke{d}}

\begin{document}

\title{Quantum-gravity-induced dual lensing and IceCube neutrinos}

\author{Giovanni Amelino-Camelia}
\affiliation{Dipartimento di Fisica, Universit\`a di Roma ``La Sapienza", P.le A. Moro 2, 00185 Roma, Italy}
\affiliation{INFN, Sez.~Roma1, P.le A. Moro 2, 00185 Roma, Italy}
\author{Leonardo Barcaroli}
\affiliation{Dipartimento di Fisica, Universit\`a di Roma ``La Sapienza", P.le A. Moro 2, 00185 Roma, Italy}
\author{Giacomo D'Amico}
\affiliation{Dipartimento di Fisica, Universit\`a di Roma ``La Sapienza", P.le A. Moro 2, 00185 Roma, Italy}
\affiliation{INFN, Sez.~Roma1, P.le A. Moro 2, 00185 Roma, Italy}
\author{Niccol\'{o} Loret}
\affiliation{Division of Theoretical Physics, Institut Ru\!$\dtagliato$\!er Bo\v{s}kovi\'{c}, Bijeni\v{c}ka cesta 54, 10000 Zagreb, Croatia}
\author{Giacomo Rosati}
\affiliation{Institute for Theoretical Physics, University of Wroc\l{}aw, Pl. Maksa Borna 9, Pl-50-204 Wroc\l{}aw, Poland}

\begin{abstract}
Momentum-space curvature, which is expected in some approaches to the
quantum-gravity problem,
can produce dual redshift,
a feature which introduces energy dependence of the travel times of ultrarelativistic particles,
and dual lensing, a feature which mainly affects the direction of observation
of particles.
In our recent arXiv:1605.00496 we explored the possibility that dual redshift might be relevant in the
analysis of IceCube neutrinos, obtaining results which are preliminarily encouraging.
Here we explore the possibility that also dual lensing might play a role
in the
analysis of IceCube neutrinos. In doing so we also investigate issues which are of broader interest,
such as the possibility of estimating the contribution by background neutrinos and some noteworthy differences
between candidate ``early neutrinos" and candidate ``late neutrinos".
\end{abstract}
\maketitle

\section{INTRODUCTION}
The possibility of an energy dependence of the travel times  from a given source to a given detector
of ultrarelativistic particles (particles whose mass is zero or is anyway negligible)
has been motivated
in several quantum-gravity-inspired models
(see {\it e.g.}
Refs.\cite{gacLRR,jacobpiran,gacsmolin,grbgac,gampul,urrutia,gacmaj,myePRL,gacGuettaPiran,steckerliberati} and references therein).
It is emerging that the most powerful perspective on this feature is that the relevant models
have their momentum space curved (see {\it e.g.}
Refs.\cite{bob,principle,kbob,Lateshift,tetradi,SpecRelLoc} and references therein),
and an effect dual to redshift produces the energy dependence of travel times.
Famously ordinary redshift due to spacetime curvature is such that particles emitted at different times with the same energy
by the same source reach the detector with different energy, and with dual redshift the curvature of momentum space
induces the feature that ultrarelativistic particles
emitted at the same time with different energy
by the same source reach the detector at different times.
It was recently realized \cite{freidelsmolin,ourlensing,transverseproceedings,stefanobiancolensing}
that typically curvature of momentum space also produces, in  addition to dual redshift,
the effect of ``dual lensing", which affects the direction of detection:
just like ordinary spacetime-curvature lensing, with dual lensing the direction by which a particle is detected
might not point to where actually the source is located.

While dual redshift has been much studied and is at this point rather well understood,
for dual lensing there have been so far only a few exploratory
 studies and several grey areas remain for its understanding. Correspondingly there is nothing much
 on model building for dual lensing in phenomenology. Aware of these challenges we nonetheless here
 explore the possibility that dual lensing might play a role in observations by the IceCube neutrino telescope.

The prediction of a neutrino emission associated with gamma ray bursts (GRBs)
is  generic within the  most widely accepted astrophysical models~\cite{fireball}.
After a few years of operation IceCube still reports \cite{icecubeUPDATEgrbnu}
no conclusive detection of GRB  neutrinos,
contradicting some influential predictions \cite{waxbig,meszabig,dafnebig,otherbig} of the GRB-neutrino observation rate by IceCube.
Of course, it may well be the case that the efficiency of neutrino production at GRBs  is much lower
than had been previously estimated~\cite{small1,small2,small3}. However, from the viewpoint of quantum-gravity/quantum-spacetime
research it is interesting to speculate that the IceCube results for GRB neutrinos might be misleading
because of the assumption that GRB neutrinos should be detected in very close temporal coincidence with
the associated $\gamma$-rays and from a direction which agrees (within errors) with the direction by which
the $\gamma$-rays are observed:
a sizeable mismatch between GRB-neutrino detection time and trigger time for the GRB could be caused by
dual redshift
(see Refs.\cite{gacLRR,jacobpiran,gacsmolin,grbgac,gampul,urrutia,gacmaj,myePRL,gacGuettaPiran,steckerliberati} and references therein)
and in presence of  dual lensing there might also be a directional mismatch.

In Ref.\cite{Ryan} we observed that allowing for dual redshift one gets a rather plausible picture in which
some of the neutrinos observed by IceCube actually are GRB neutrinos.
We here
 explore the possibility that also dual lensing might play a role
in the
analysis of IceCube neutrinos. In doing so we investigate issues which are also relevant for more  refined
analyses of dual redshift,
such as the possibility of estimating the contribution by background neutrinos and some noteworthy differences
between candidate ``early neutrinos" and candidate ``late neutrinos".

\section{Dual redshift and its possible interplay with dual lensing}
Dual redshift has been discussed extensively (with or without reference to the role played by momentum-space curvature)
in the context of
some much-studied models of
spacetime quantization (see, {\it e.g.}, \cite{jacobpiran,gacsmolin,gacLRR,grbgac,gampul,urrutia,gacmaj,myePRL} and references therein).
For the type of data analyses we are interested in, dual redshift has the implication that
 the time needed for a ultrarelativistic particle\footnote{Of course the only regime of particle propagation
 that is relevant for this manuscript is the ultrarelativistic regime, since photons have no mass and
 for the neutrinos we are contemplating (energy of
 tens or hundreds of TeVs) the mass is completely negligible.}
to travel from a given source to a given detector receives a quantum-spacetime correction, here denoted with $\Delta t$.
We here focus, as in Ref.\cite{Ryan}, on the class of scenarios whose predictions for $\Delta t$ can all be described,
for corresponding choices of the parameters $\eta$ and $\delta$, in terms of the formula
(working in units with the speed-of-light scale ``$c$" set to 1)
\begin{equation}
\Delta t = \eta \frac{E}{M_{P}} D(z) \pm \delta \frac{E}{M_{P}} D(z) \, .
\label{main}
\end{equation}
Here the redshift- ($z$-)dependent  $D(z)$ carries the information on the distance between source and detector, and it factors
in the interplay between quantum-spacetime effects and the curvature of spacetime.
As usually done in the relevant literature \cite{jacobpiran,gacsmolin,gacLRR} we take for $D(z)$ the following form:\footnote{The interplay between quantum-spacetime effects and curvature of spacetime is still a lively subject of investigation, and, while (\ref{dz})
is by far the most studied scenario, some alternatives to (\ref{dz}) are also under consideration \cite{dsrfrw}.}
\begin{equation}
D(z) = \int_0^z d\zeta \frac{(1+\zeta)}{H_0\sqrt{\Omega_\Lambda + (1+\zeta)^3 \Omega_m}}  \, ,
\label{dz}
\end{equation}
where $\Omega_\Lambda$, $H_0$ and $\Omega_0$ denote, as usual,
respectively the cosmological constant, the Hubble parameter and the matter fraction, for which we take the values given in Ref.\cite{PlanckCosmPar}.
With $M_{P}$ we denote the Planck scale ($\simeq 1.2\,\cdotp 10^{28}eV$) while
the values of the  parameters $\eta$ and $\delta$ in (\ref{main})
characterize the specific scenario one intends to study. In particular, in (\ref{main}) we used the notation ``$\pm \delta$"
to reflect the fact that $\delta$ parametrizes the size of quantum-uncertainty (fuzziness) effects. Instead the parameter $\eta$
characterizes systematic effects: for example in our conventions for positive $\eta$ and $\delta =0$ a high-energy neutrino
is detected systematically after a low-energy neutrino (if the two neutrinos are emitted simultaneously).

The parameters $\eta$ and $\delta$ are to be determined experimentally. When non-vanishing,
they are expected to take values somewhere in a neighborhood of 1, but values as large as $10^3$ are plausible if the solution
to the quantum-gravity problem is somehow connected with the unification of non-gravitational forces \cite{gacLRR,wilczek} while values
significantly smaller than 1 find support in some renormalization-group arguments (see, {\it e.g.}, Ref.\cite{hsuHIGGSES}).
In general, $\eta$ and $\delta$ can take different values for different particles  \cite{gacLRR,myePRL,mattiLRR,szabo1}
and the
realm of scenarios for the particle dependence of the effects is rather wide. In particular, one should allow for a dependence of $\eta$ and $\delta$ on the helicity~\cite{gacLRR,mattiLRR} of the neutrino, with the net result of a neutrino phenomenology governed by four parameters, $\eta_+$, $\delta_+$, $\eta_-$, $\delta_-$.
Presently for photons the limits on $\eta$ and $\delta$ are at the level of $|\eta| \lesssim 1$ and $\delta \lesssim 1$ \cite{fermiNATURE,gacNATUREPHYSICS2015}, but for neutrinos we are still several orders of magnitude below 1 \cite{steckerliberati,gacLRR}.

Much less is known about dual lensing. The few toy models which have been shown to exhibit it \cite{freidelsmolin,ourlensing,transverseproceedings,stefanobiancolensing} are probably
not representative of the variety of possibilities one should consider for dual lensing in the quantum-gravity realm.
We shall here adopt an exploratory attitude. The key point for us is just to explore the issue
that GRB neutrinos might not be identified as such both because of time-of-arrival effects (dual redshift) and
directional effects (dual lensing).

\section{Summary of our recent analysis for pure dual redshift}

We set the stage for our investigations of dual lensing by revisiting the most significant points of the
analysis we recently reported in Ref.\cite{Ryan}, tentatively assuming pure dual redshift.

It is convenient to introduce a ``distance-rescaled time delay" $\Delta t^*$ defined as
\begin{equation}
\Delta t^* \equiv \Delta t \frac{D(1)}{D(z)}
\label{tstar}
\end{equation}
so that (\ref{main}) can be rewritten as
\begin{equation}
\Delta t^* = \eta \frac{E}{M_{P}} D(1) \pm \delta \frac{E}{M_{P}} D(1) \, .
\label{maintwo}
\end{equation}
This reformulation of (\ref{main}) allows to describe the relevant quantum-spacetime effects,
which in general depend both on redshift and energy, as effects that depend exclusively on energy,
through the simple expedient of focusing on the relationship between $\Delta t$ and energy
when the redshift has a certain chosen value, which in particular we chose to be $z=1$.
If one measures a certain $\Delta t$ for a candidate GRB neutrino and the redshift $z$ of
the relevant GRB is well known, then one gets a firm determination of $\Delta t^*$ by simply rescaling
the measured $\Delta t$ by the factor $D(1)/D(z)$. And even when the redshift
of the relevant GRB is not known accurately one will be able to convert a measured $\Delta t$
into a determined $\Delta t^*$ with accuracy governed by how much one is able to still assume
about the redshift of the relevant GRB. In particular, even just the information on whether a GRB is long
or short can be converted into at least a very rough estimate of redshift.

In order to select some GRB-neutrino candidates
we need a temporal window (how large can the $\Delta t$ be in order for us to consider a
IceCube event as a potential GRB-neutrino candidate) and we need criteria of directional selection (how well the directions estimated
for the IceCube event and for the GRB should agree in order for us to consider that IceCube event as a potential GRB-neutrino candidate).
We focus \cite{Ryan} on neutrinos with energies
between 60 TeV and 500 TeV, allowing
for a temporal window of 3 days.  We based \cite{Ryan} our directional criteria for the selection of GRB-neutrino candidates on
the signal direction PDF depending on the space angle difference between GRB and neutrino: $P(\nu,GRB)=(2\pi\sigma^2)^{-1}\exp(-\frac{|\vec{x}_{\nu}-\vec{x}_{GRB}|^2}{2\sigma^2})$, a two dimensional circular Gaussian whose standard deviation is
\begin{equation}\label{sigmas}
\sigma=\sqrt{\sigma_{GRB}^2+\sigma_{\nu}^2} \, ,
\end{equation}
where of course $\sigma_{GRB}$ and $\sigma_{\nu}$ denote respectively the
uncertainties in  the direction of observation of the GRB and of the neutrino.
We then request that a GRB-neutrino candidate should be such that
 the pair composed by the neutrino and the GRB is at angular distance compatible within a 2$\sigma$ region.

A key observation of our Ref.\cite{Ryan}  is that whenever $\eta$ and/or $\delta$ do not vanish one should expect
on the basis of (\ref{maintwo}) a correlation between the $|\Delta t^*|$ and the energy of the candidate GRB neutrinos.

Our data set is for four years of operation of IceCube \cite{IceCube},
  from June 2010 to May 2014.
Since the determination of the energy of the neutrino plays such a  crucial role in our analysis
we include only IceCube ``shower events" (for ``track events" the reconstruction of the neutrino energy is far more problematic  and less reliable \cite{IceCubeBackground,TRACKnogood1}).
We have 21 such events within our 60-500 TeV energy window, and we find that 9 of them
fit our requirements for candidate GRB neutrinos.
The properties of these 9 candidates that are most relevant for our analysis are summarized in Table 1 and Figure 1.

\begin{table}[htbp]
\centering
{\def\arraystretch{0.5}\tabcolsep=3pt
\begin{tabular}{c|c|c|l|r|c|c}
\hline
$\,$                   &    \!\!\!   E \!\!\!\! [TeV]      \!\!\!        & GRB              & z           & $\Delta t^*$ [s]      & $\,$  & $\,$   \\\hline \hline
IC9                          &                 63.2                 & 110503A    & 1.613        & 50227       &  $\diamond$   &     *    \\\hline
IC19                         &                71.5                  & 111229A    & 1.3805        & 53512        & $\,$   &     *  \\\hline
\multirow{3}{*}{IC42} & \multirow{3}{*}{76.3}    & 131117A      & 4.042    & 5620       &         & $\,$     \\
&                                       & 131118A      & 1.497 *   &  -98694      & $\,$     &     *    \\
&                                       & 131119A      & \,\,\,\,\, ? &  -146475           &  $\diamond$  &   $\,$      \\
\hline
IC11                          &                88.4                 & 110531A    & 1.497   *     & 124338       & $\,$  &     *     \\\hline
IC12                         &                104.1                 & 110625B    & 1.497   *     & 108061      & $\,$   &     *    \\\hline
\multirow{3}{*}{IC2} & \multirow{3}{*}{117.0}    & 100604A      & \,\,\,\,\, ?  & 10372           &      & $\,$     \\
                                &                                       & 100605A      & 1.497 *    &  -75921      & $\,$   &     * \\
                                &                                       & 100606A      & \,\,\,\,\, ?  &  -135456     &          & $\,$           \\\hline

IC40                          &             157.3                 & 130730A    & 1.497   *     & -120641       & $\diamond$   &     *    \\\hline
\multirow{2}{*}{IC26}
& \multirow{2}{*}{210.0}   & 120219A      & 1.497  *  &  153815    & $\,$     &     *    \\
&                          & 120224B     & \,\,\,\,\, ? &  -117619     &       &  $\,$  \\
\hline
IC33                          &             384.7                 & 121023A    & \,\, 0.6      *     &  -289371       & $\,$   &     *   \\\hline
\end{tabular}
}
\caption{Among the 21 ``shower neutrinos" with energy between 60 and 500 TeV observed by IceCube between June 2010 and May 2014
only 9 fit our directional and temporal criteria for GRB-neutrino candidates, and yet for 3 of them there is more than one GRB
to be considered when pairing up neutrinos and GRBs. The last column highlights with an asterisk the 9 GRB-neutrino candidates ultimately selected in our Ref.\cite{Ryan}
by our additional criterion of maximal correlation. Additionally here we highlight with a $\diamond$ the 3 cases such that
the pair composed by the neutrino and the GRB is at angular distance compatible within a $\sigma$ region
(all other candidates are compatible within a 2$\sigma$ region).
Also shown in table are the values of redshift attributed to the relevant GRBs:
the redshift is known only
for GRB110503A, GRB111229A and GRB131117A.
GRB111229A and GRB110503A are long GRBs and we assume that the average of their redshifts (1.497) could be a reasonably good estimate
of the redshifts of the other long GRBs relevant for our 9 GRB-neutrino candidates. These are the 6 estimated values of redshift
$z=1.497^*$, the asterisk reminding that it is a ``best guess" value. For analogous reasons we place an asterisk close to the value of 0.6 which is our best guess for the redshift of the only short GRB
in our sample. The first column lists the ``names" given by IceCube to the neutrinos that end up being relevant for our analysis.}
\label{table1}
\end{table}

\begin{figure}[h!]

\includegraphics[scale=0.3]{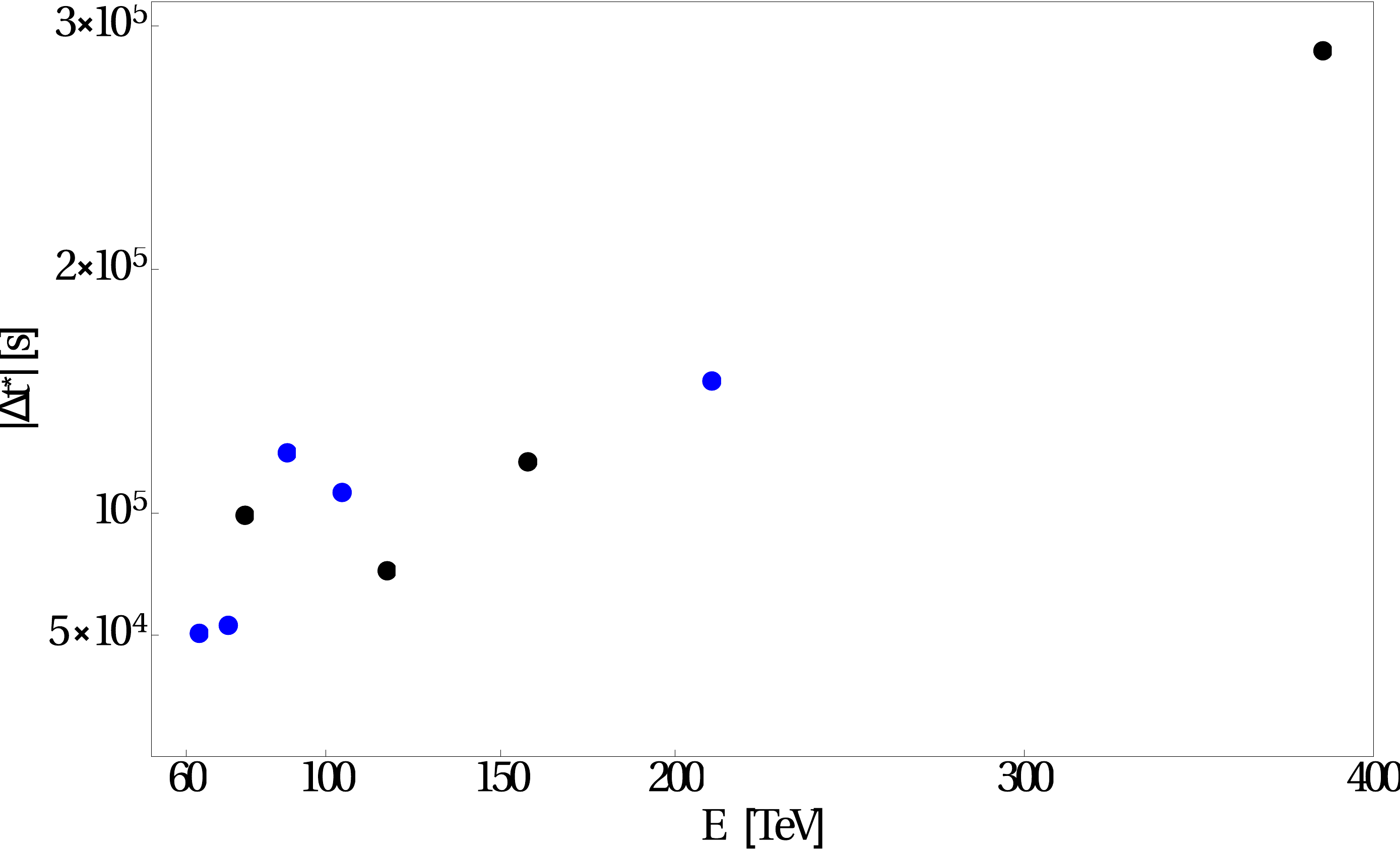}
\caption{Points here in figure correspond to the 9 GRB-neutrino candidates highlighted with an asterisk in the last column of Table 1. Blue points correspond to ``late neutrinos" ($\Delta t^*>0$),
while black points correspond to ``early neutrinos" ($\Delta t^*<0$).}
\end{figure}

As visible in Table 1,
for some IceCube events our selection
criteria produce multiple GRB-neutrino candidates.
In Ref.\cite{Ryan} we handled this issue of multiple candidates
by focusing on the case
 that provides the highest correlation.

Another issue reflected by Table 1 comes from the fact that for only 3 of the GRBs involved in this analysis the redshift is known.  We must handle only one short GRB of unknown redshift, and we assume for it a redshift of 0.6,
which is a rather reasonable rough estimate for a short GRB.
For some of our long GRBs we do have a redshift determination and we argued in Ref.\cite{Ryan}
that consistently with the hypothesis here being tested
one should use those known values of redshift for obtaining at least a rough estimate of the redshift of long GRBs for which
the redshift is unknown. This is illustrated by the 9 GRB-neutrino candidates marked by an asterisk in table 1:
those 9 candidates include 8 long GRBs, 2 of which have known redshift, and we assign to the other 6 long GRBs
the average ${\bar{z}}$  of those two values of redshift (${\bar{z}}=1.497$).

The correlation between $|\Delta t^*|$ and energy
 for the 9 GRB-neutrino candidates highlighted in Fig.1
 is of 0.951.
This is a strikingly high value of correlation but in itself
does not provide what is evidently the most interesting quantity here of interest,
which must be some sort of ``false alarm probability": how likely it would be to have accidentally data with such
good agreement with the expectations of the quantum-spacetime models here contemplated?
We proposed in Ref.\cite{Ryan} that one needs to estimate how often
a sample composed exclusively of background neutrinos\footnote{Consistently with the objectives of our analysis we consider as ``background neutrinos" all
 neutrinos that are unrelated to a GRB, neutrinos of atmospheric or other astrophysical origin which end up being selected
as GRB-neutrino candidates just because accidentally their time of detection and angular direction happen to fit
our selection criteria.} would produce accidentally
9 or more GRB-neutrino candidates with correlation
comparable to (or greater than) those we found in data.
We did this by performing $10^5$ randomizations of the times of detection
of the 21 IceCube neutrinos relevant for our analysis, keeping their energies and directions fixed,
and for each of these time randomizations we redo the analysis\footnote{In particular for any given realization of the fictitious GRB-neutrino candidates we identify those of known redshift and use them to estimate the ``typical fictitious GRB-neutrino redshift", then attributed to
those candidates of unknown redshift (procedure done separately for long and for short GRBs). When in the given
  realization of the fictitious GRB-neutrino candidates there is no long (short) GRB of known redshift we attribute to all of them
a redshift of 1.497 (0.6).} just as if they were real data.
Our observable is a time-energy correlation and by randomizing the
times we get a robust estimate of how easy (or how hard) it is for a sample composed exclusively
of background neutrinos to produce accidentally
a certain correlation result.
In the analysis of these fictitious data obtained by randomizing the detection times of the neutrinos
we handle cases with neutrinos for which there is more than one possible GRB partner by maximizing the correlation,
in the sense already discussed above for the true data. We ask how often this time-randomization
procedure produces 9 or more GRB-neutrino candidates with correlation
$\geq 0.951$, and remarkably we found that this happens only in 0.03$\%$ of cases.

\section{Background neutrinos and dual lensing}

In our previous paper (Ref.\cite{Ryan}) the directional analysis was limited to the requirement
that the pair composed by the neutrino and the GRB should be at angular distance compatible within a 2$\sigma$ region.
It is however interesting that (as one can easily check from Ref.\cite{IceCube} and from the useful tools provided by the IceCube collaboration\footnote{See at http://grbweb.icecube.wisc.edu/})
for only 2 of our 9 GRB-neutrino candidates the pair composed by the neutrino is at angular distance compatible within one $\sigma$.
The 2 relevant GRB-neutrino candidates are IC9/GRB110503A and IC40/GRB130730A.
It is interesting to contemplate this fact taking as working assumption, just for the sake of this exercise,
that the model we are considering is correct, and therefore there are GRB neutrinos of the type here considered.
 If our 9 GRB-neutrino candidates were all ``signal" neutrinos, and if the directional uncertainties were correctly estimated,
we should expect 6 or 7 candidates at one sigma (since there are 9 candidates at 2 sigma).
Finding 2 instead of 6 or 7 could be significant.

In part this can be blamed on our selection criteria: for cases in which for one neutrino there was more than one GRB
possibly associated to it we handled the multiplicity by selecting the maximum-correlation option, as explained above.
We could have used the maximum correlation criterion only at a lower level of selection, giving priority
instead to
GRB-neutrino candidates such that the pair composed by the neutrino and the GRB is at angular distance compatible within one $\sigma$.
However, if this alternative criterion had been adopted only 1 of our 9 candidates would be affected:
we would select IC42/GRB131119A instead of IC42/GRB131118A, with the end result that 3 out of the 9 candidates
would have direction compatible at one  sigma.

Having 3 at one sigma out of 9 at two sigmas is still not very satisfactory. However, there are at least two
possible explanations:

(i) We surely have some background neutrinos among our candidates and if, say, 4 of the 9 are background then one would be in
the situation of having  3 out of 5 signal neutrinos with direction acceptable at one sigma,
which is satisfactory.

(ii) Dual lensing could be responsible:
the low number of GRB-neutrino candidates ``at one sigma" would be expected
if there is a sizable mismatch between the apparent direction of observation of the neutrino and
the actual direction pointing to its source.

In this section we mainly explore the first hypothesis, (i), while the next section is focused on the second hypothesis.

\subsection{Estimating background}

In light of the observations reported at the beginning of this section it is of paramount importance to have at least a rough estimate of how many
of our 9 GRB-neutrino candidates should be expected to be background.
For this purpose it is useful to notice that out of the 21 neutrinos in our sample only 9 turned out to fit
our requirements for GRB-neutrino
candidates. We have therefore 12 neutrinos which are background in both of the scenarios we are comparing: if all 21 neutrinos are background of course also those 12 are background, and even if some of the neutrinos are GRB neutrinos of the type
we are contemplating we find that those 12 still must be interpreted as background.

We can therefore ask how likely it would have been for one or more of those 12 neutrinos to accidentally appear
to be GRB neutrinos of the type we are looking for. This can be estimated by randomizing the times of those 12 neutrinos.
Of course, if, say, it is likely that 4 of those neutrinos could appear as GRB neutrinos, we will assume
that a proportionate number of our 9 GRB-neutrino candidates are background. It is evidently an estimate to
be performed by self-consistence: one starts with 12 neutrinos which are surely background, but then one is led to increase
the estimated number of background neutrinos, since the analysis itself suggests that also some of our 9 GRB-neutrino
candidates actually are background.

Among the results obtained applying this logic of analysis, randomizing the times of the 12 sure-background neutrinos,
we stress in particular that there is a probability of 66$\%$ that 3, 4 or 5 of our 9 GRB-neutrino candidates
are background (19$\%$ that 3 of them are background, 26$\%$ that 4 of them are background, and 21$\%$ that 5 of them are background).
This by itself renders already less surprising the fact that we have only 3 GRB-neutrino candidates
at one  sigma, with 6 more at two sigmas: it may well be that the 3 at one sigma are all signal, while, say, 4 of the remaining 6
are background.

\subsection{Overachieving background neutrinos}
The fact that it is very
likely that 3, 4 or 5 of our 9 GRB-neutrino candidates
actually are background renders the ``directional story" of our analysis less surprising,
but in turn brings up some questions concerning the very high correlation we found in Ref.\cite{Ryan}.
We have a monstrous correlation of 0.951 for our 9 GRB-neutrino candidates even though very likely 4 or 5 of them
are background!
In the corresponding sense our 9 candidates overachieved.

It is interesting to estimate at least roughly by how much our 9-plet overachieved. Let us do that by taking as reference
the case that 4 among our 9 GRB-neutrino candidates are background.
We can then exploratively assume that the 5-plet of ``true" GRB neutrinos is the maximum-correlation 5-plet among the 5-plets obtainable from our 9 candidates.
 These are IC9/GRB110503A, IC19/GRB111229A, IC40/GRB130730A, IC26/GRB120219A, IC33/GRB121023A.
Accordingly we would have that the remaining four,
IC11/GRB110531A, IC42/GRB131118A, IC12/GRB110625B, IC2/GRB100605A, are background. But these remaining four still contribute
rather strongly to the correlation: the correlation of 0.9996 of the maximum-correlation 5-plet is only decreased to 0.951
when we include all 9 candidates. Within the assumptions we are making this means that the 4 background neutrino ``overachieved",
{\it i.e.} they did not behave like standard background neutrinos, but rather, accidentally, looked like signal neutrinos.
We can quantify this overachievement by randomizing the times
of detection of IC11/GRB110531A, IC42/GRB131118A, IC12/GRB110625B, IC2/GRB100605A (keeping the times
 of detection of the other 5 neutrinos fixed) and seeing what is the expected
value of correlation. This value is 0.903.
Interestingly if we take as reference the value of correlation of 0.903
the corresponding false alarm probability, computed just as prescribed in our Ref.\cite{Ryan},
is of 0.21$\%$ (still very small but significantly higher than the false alarm probability of 0.03$\%$ one estimates
ignoring this issue of the overachieving background neutrinos).

When taking as working assumption a certain number of background neutrinos it makes sense to  compute
a false alarm probability defined in a slightly different way from the one introduced in  our Ref.\cite{Ryan}.
We introduce this notion focusing again, for illustrative purposes, on the case in which one takes
as working assumption that 4 among our 9 GRB-neutrino candidates are background, then computing for the true data
the maximum value of correlation obtainable by considering 5 out of our 9 candidates (the highest value
of correlation found among the 126 possible 5-plets of candidates obtainable from our total of 9 candidates).
We can define a false alarm probability based on how frequently simulated data, obtained by randomizing the times
of detection of the 21 neutrinos in our sample, include a 5-plet of candidates with correlation greater or equal to
the one found for the best 5-plet in the real data (so, if, say, a given time randomization produces 11 candidates one
assigns to the randomization a value of correlation given by the highest correlation found by considering all possible choices
 of 5 out of the 11
candidates). We find that this false alarm probability is of 0.16$\%$.

\section{Probing the possible presence of dual lensing}
In the previous section we showed that the  ``directional story" of our analysis is not so surprising in light of
a plausible estimate of the role played by background neutrinos.
Since probably 4 or 5 of our 9 candidates are background neutrinos that only accidentally we selected as candidates, it
 is not surprising that only 2 (3 with another criterion discussed above) candidates are ``at one sigma"
 among the total of 9 candidates
that we have ``at two sigmas". This renders less compelling the hypothesis that dual lensing might have payed a role, but
of course, one may nonetheless explore the possible
role of dual lensing. We shall do this knowing that, as the interested reader will easily realize,
analyses such as ours would be affected tangibly
by dual lensing only if  rather large directional mismatches are produced, at least as large as a few degrees.
We find it hard to believe that a quantum-gravity effect could be this large, but of course we still
rely on data rather than prejudice to investigate the issue.

The magnitude of dual lensing will likely depend on the energy of the particle \cite{freidelsmolin,ourlensing,transverseproceedings,stefanobiancolensing};
however, for this exploratory study we shall be satisfied with a rudimentary and limited
description of dual lensing: we shall simply assume that the directional uncertainty of the neutrinos, for
which we used above the notation $\sigma_\nu$,
receives an additional energy-independent contribution $\sigma_{d.l.}$,
$$\sigma_\nu \rightarrow \sigma'_\nu = \sigma_\nu + \sigma_{d.l.}$$
We are therefore adopting a rudimentary description of dual lensing which is energy independent and is of ``fuzzy type" (non-systematic),
so that (\ref{sigmas}) is replaced by
\begin{equation}\label{sigmasnew}
\sigma'=\sqrt{\sigma_{GRB}^2+(\sigma_{\nu}+ \sigma_{d.l.})^2} \, ,
\end{equation}

We only consider a few values of $\sigma_{d.l.}$, specifically 5, 10, 15 and 20 degrees. With more data it would
make sense to probe $\sigma_{d.l.}$ more finely, but in the present situation this is evidently sufficient.
What we are looking for is establishing whether or not the additional GRB-neutrino candidates picked up by allowing
for $\sigma_{d.l.}$ manifest any connection with the 9 GRB-neutrino candidates we had with our original analysis
of Ref.\cite{Ryan}.
We select candidates just as discussed in Sec.III, but now replacing the $\sigma$ of (\ref{sigmas}) with
the  $\sigma'$ of (\ref{sigmasnew}), for the few mentioned values  of $\sigma_{d.l.}$.
We find that for $\sigma_{d.l.} =5$degrees one picks up 2 additional GRB-neutrino candidates in addition
to the 9 candidates we already had for $\sigma_{d.l.} =0$. These two ``dual-lensing candidates" selected with the criteria
of Sec.III (but with $\sigma_{d.l.} =5$degrees) are IC39/GRB130707A and
IC46/GRB140129C. The correlation of  our original 9 candidates was 0.951 and with addition of
IC39/GRB130707A and
IC46/GRB140129C the correlation goes down to 0.830. The fact that the correlation goes down does not in itself
provide an indication against dual lensing, since 0.830 is still a very high value. The relevant issue for
assessing the ``performance of dual lensing" concerns whether these 2 additional candidates ``look like background" or rather
appear to be in reasonably good agreement with the 9 candidates we already had. For luck of a better name we shall label this
as the ``variation probability". While it is intuitively clear what one intends to characterize with such a variation probability,
 there are, as we shall see, at least a couple of possibility for its definition that one should consider.
  Let us start by estimating a variation probability by randomizing the times of the 12 IceCube neutrinos not involved in our original 9
candidates, and seeing how frequently one picks up extra candidates, by allowing for $\sigma_{d.l.}$ of 5 degrees,
such that the overall correlation is of 0.831 or higher. We find that this ``variation probability" is 35$\%$,
providing an indication which is (however mildly) favorable for the dual-lensing hypothesis with $\sigma_{d.l.} =5$degrees:
if the 12 relevant neutrinos are all background then in 65$\%$ of cases
one would expect to find a correlation lower than 0.831.
An alternative definition of the variation probability could fix the number of ``dual-lensing candidates" found
in simulations: one would
randomize the times of the 12 IceCube neutrinos not involved in our original 9
candidates, and focus on cases when such randomizations produce a number of dual-lensing candidates
equal to the number of dual-lensing candidates found on true data.
We  shall label this second notion of variation probability as the ``fixed-number variation probability",
and we find that for $\sigma_{d.l.}$ of 5 degrees, this fixed-number variation probability is of
69$\%$,
which is evidently less encouraging for dual lensing.
The main reason for contemplating the possibility of
fixing the number of dual-lensing candidates in simulations to be equal to the number of
dual-lensing candidates
found in true data is that we are focusing on variations of the correlation and the typical size of such variations
depends of course on how many additional candidates contribute. Starting with 9 candidates, if we only add, say, one more candidate
the variation typically will be small, significantly smaller than what one typically should find for cases with, say, 5 additional candidates.

Next we consider the case $\sigma_{d.l.} =10$degrees, finding that in that case one picks up 3 additional GRB-neutrino candidates in addition
to the 9 candidates we already had for $\sigma_{d.l.} =0$, one more, which is IC22/GRB120114B, in addition to the 2 already
found for $\sigma_{d.l.} =5$degrees. For the total of 12 GRB-neutrino candidates selected for $\sigma_{d.l.} =10$degrees
one has correlation of 0.770. Moreover we find a variation probability of 45$\%$  by randomizing the times of the 12 IceCube neutrinos not involved in our original 9
candidates, and seeing how frequently one picks up extra candidates, by allowing for $\sigma_{d.l.}$ of 10 degrees,
such that the overall correlation is of 0.770 or higher.
A variation probability of 45$\%$ is not at all  encouraging: it just  means that overall the extra candidates picked up
for $\sigma_{d.l.} =10$degrees behaves just like typically one would expect pure background to behave.
This is confirmed by computing the fixed-number false alarm probability, which turns out to be of 73$\%$.

For $\sigma_{d.l.} =15$degrees and $\sigma_{d.l.} =20$degrees one gets a picture very similar to what
we just reported for $\sigma_{d.l.} =10$degrees, not favorable to dual lensing.
For $\sigma_{d.l.} =15$degrees
one picks up one more candidate, which is
IC30/GRB120709A, and the candidate IC39/GRB130705A, selected already at $\sigma_{d.l.} =5$degrees, is replaced\footnote{As explained above and in Ref.\cite{Ryan}, we are consistently using a criterion
such that, when a given neutrino has available more than one GRB partner, the GRB-neutrino candidate is taken to
be the one that maximizes the correlation. For the neutrino IC39 up to $\sigma_{d.l.} =10$degrees
one has only one possible GRB partner, which is GRB130707A, but at $\sigma_{d.l.} =10$degrees one has that also GRB130705A
becomes directionally compatible with IC39, and actually the candidate IC39/GRB130705A leads to higher correlation
than IC39/GRB130707A.} by IC39/GRB130707A;
the correlation for the total of 13 candidates is 0.767, giving a variation probability of 44$\%$
(and a fixed-number variation probability of 72$\%$).
At $\sigma_{d.l.}$ of 20 degrees one picks up 4 more candidates, for a total of 17  (the 9 candidates we already had
for $\sigma_{d.l.} =0$, the 4 other candidates we had already picked up going up to $\sigma_{d.l.} =15$degrees, plus 4 more
candidates). The correlation for this 17 candidates is 0.716,
giving a variation probability of 62$\%$
(and a fixed-number variation probability of  67$\%$).

\section{Differentiating between early neutrinos and late neutrinos }
We have so far, like in our previous Ref.\cite{Ryan},
considered values of correlation between energy and the absolute value of $\Delta t^*$, thereby having
 a situation such that both early neutrinos (neutrinos observed before  the relevant GRB)
and late neutrinos (neutrinos observed after the relevant GRB) contribute on the same footing to the same correlation study.
On the theory side this is relevant for scenarios in  which, in the sense of our
Eq.(\ref{main}) (and of comments to Eq.(\ref{main}) which we offered in Sec.II), one has that\footnote{Evidently $\delta_+$ and $\delta_-$ automatically do not discriminate between early and late neutrinos since
they govern uncertainty-type effects.} $\eta_+ \simeq - \eta_-$.
A posteriori this found some motivation in the content of our data set, since we found about an equal number
of early neutrinos and late neutrinos. However, we here argued that quite a few of our GRB-neutrino candidates
(as many as 4 or 5 out of 9) are likely background and if these were mostly of one type  (say, all late neutrinos)
our perspective on the data could change significantly. This motivates us to also consider separately early neutrinos
and late neutrinos.

In the next subsection we do this adopting the strategy of analysis of our previous Ref.\cite{Ryan},
which in particular required that
 the pair composed by the neutrino and the GRB should be at angular distance compatible within a 2$\sigma$ region
 ($\sigma_{d.l.}=0$).
Then we also perform the same analysis making room for dual lensing,
requiring that
 the pair composed by the neutrino and the GRB should be at angular distance compatible within a $2 \sigma'$ region,
 for values of $\sigma_{d.l.}$ of 5, 10, 15 and 20.

\subsection{Early and late neutrinos without dual lensing}
So let us start by going back to $\sigma_{d.l.}=0$, as in our previous Ref.\cite{Ryan}, but now looking separately at early neutrinos
and late neutrinos.
For the 21 IceCube shower neutrinos of energy between 60 and 500 GeV we ask that a potential GRB partner for a late (early) neutrino
should be observed
up to 3 days earlier (later) than the neutrino, and that
 the pair composed by the neutrino and the GRB should be at angular distance compatible within a 2$\sigma$ region ($\sigma_{d.l.}=0$).

We find 5 ``early GRB-neutrino candidates", which we list in Table 2. Of course this is a subset of the content of Table 1.
Figure 2 illustrates these findings.

\begin{table}[htbp]
\centering
{\def\arraystretch{0.3}\tabcolsep=3pt
\begin{tabular}{c|c|c|l|r|c}
\hline
$\,$                   &    \!\!\!   E \!\!\!\! [TeV]      \!\!\!        & GRB              & z           & $\Delta t^*$ [s]      & $\,$   \\\hline \hline
\multirow{3}{*}{IC42} & \multirow{3}{*}{76.3}

                                         & 131118A      & 1.497  *  &  -98694       &     *      \\
&                                       & 131119A      & \,\,\,\,\, ? &  -146475     &             \\
\hline
\multirow{3}{*}{IC2} & \multirow{3}{*}{117.0}
                                                                         & 100605A      & 1.497 *    &  -75921       &     * \\
                                &                                       & 100606A      & \,\,\,\,\, ?  &  -135456     &                   \\\hline

IC40                          &             157.3                 & 130730A    & 1.497   *     & -120641       &       *  \\\hline
IC26  & 210.0           & 120224B     & 1.497   * &  -117619    &         *  \\
\hline
IC33                          &             384.7                 & 121023A    & \,\, 0.6      *     &  -289371       &    *     \\\hline
\end{tabular}
}
\caption{This table uses the same conventions as Table 1, but includes exclusively early neutrinos.}
\label{table2}
\end{table}

\begin{figure}[h!]

\includegraphics[scale=0.3]{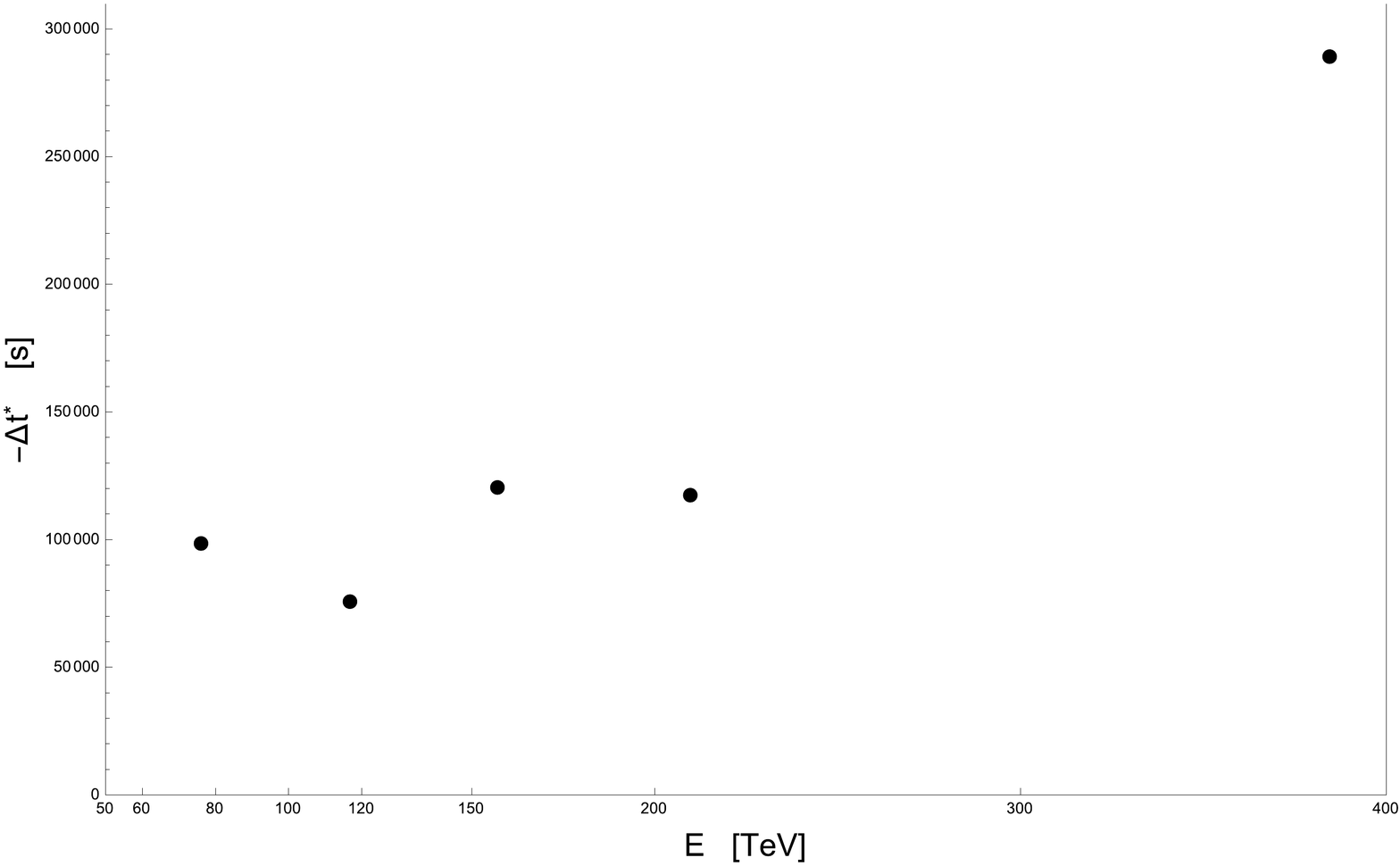}
\caption{Points here in figure correspond to the 5 ``early GRB-neutrino candidates" highlighted with an asterisk in the last column of Table 2.}
\end{figure}

For these 5 early neutrinos the correlation between energy and\footnote{For early neutrinos it is natural
to consider the correlation between energy and $- \Delta t^*$, since by changing the overall sign of $\Delta t^*$
for early neutrinos one gets results more readily comparable to the ones obtained for the correlation
between energy and $\Delta t^*$ for late neutrinos.} $- \Delta t^*$ takes the very high value of 0.945.
The resulting false alarm probability (computed as in our previous Ref.\cite{Ryan}
but now focusing only on early neutrinos)
is correspondingly very low, a false alarm probability of only  0.56$\%$.

We of course redo the same exercise for late neutrinos.
We find 7 ``late GRB-neutrino candidates", which we list in Table 3, while
Figure 3 illustrates these findings.

\begin{table}[htbp]
\centering
{\def\arraystretch{0.3}\tabcolsep=3pt
\begin{tabular}{c|c|c|l|r|c}
\hline
$\,$                   &    \!\!\!   E \!\!\!\! [TeV]      \!\!\!        & GRB              & z           & $\Delta t^*$ [s]      & $\,$   \\\hline \hline
IC9                          &                 63.2                 & 110503A    & 1.613        & 50227       &  *       \\\hline
IC19                         &                71.5                  & 111229A    & 1.3805        & 53512       &   *    \\\hline
IC42                          &              76.3                   & 131117A      & 4.042        & 5620              &  *          \\
\hline
IC11                          &                88.4                 & 110531A    & 2.345  *     & 124338       &  *       \\\hline
IC12                         &                104.1                 & 110625B    & 2.345   *     & 108061       &   *      \\\hline
IC2                           &               117.0                   & 100604A   &  2.345 *      & 10372       &    *        \\ \hline

IC26                         &              210.0                    & 120219A   & 2.345 *       &  153815     &   *        \\

\hline
\end{tabular}
}
\caption{This table uses the same conventions as Table 1, but includes exclusively late neutrinos.}
\label{table3}
\end{table}

\begin{figure}[h!]

\includegraphics[scale=0.3]{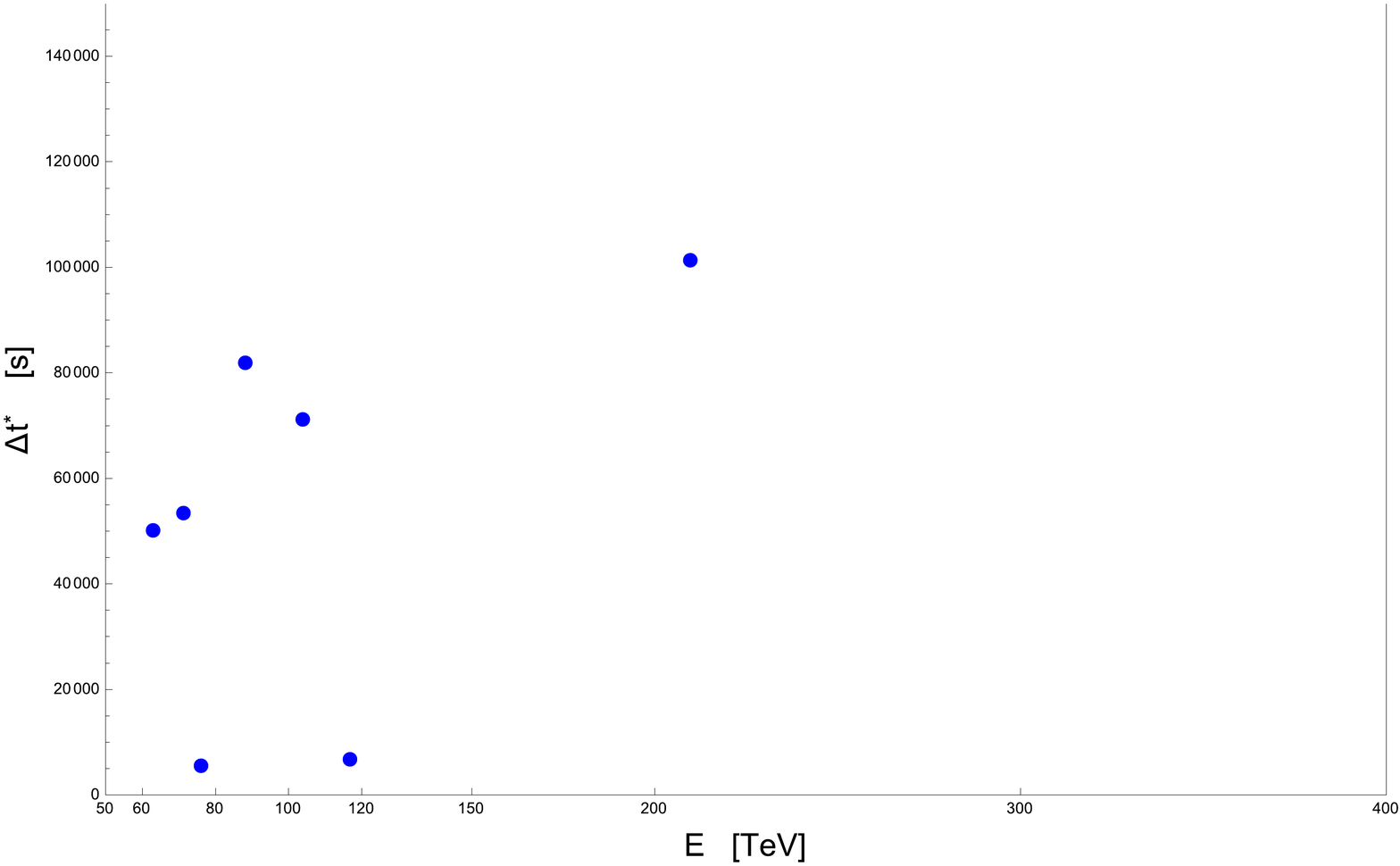}
\caption{Points here in figure correspond to the 7 ``late GRB-neutrino candidates" highlighted with an asterisk in the last column of Table 3.}
\end{figure}

For these 7 late neutrinos the correlation between energy and $\Delta t^*$ takes the value of 0.502,
which is still rather high, but significantly smaller than the one found for early neutrinos.
Of course, in light of the estimates concerning background neutrinos offered above,
this does not necessarily imply that the case for early neutrino is stronger than
than for late neutrinos (as many as 3 or 4 of these 7 late neutrinos could be coherently interpreted
as background, even if one took as working assumption the presence of dual redshift).
Setting aside the possible role of background neutrinos, one can observe
that the resulting false alarm probability
(computed as in our previous Ref.\cite{Ryan}
but now focusing only on late neutrinos)
is of 2.6$\%$, indeed rather small but not as small as for the early-neutrino case.

We should also stress than when differentiating between early and late neutrinos one  does not necessarily have to focus
on one or the other, in the sense that some valuable information could be found also by searching for both early and late neutrinos
while keeping track of their difference. This could be particularly valuable
if $\eta_+$ and $\eta_-$ have opposite sign and significantly different size (the effect for late neutrinos having different
size from the effect for early neutrinos).
In such a case one could compute separately the correlation found in late neutrinos, which one could denote
 by $\rho_+$, and the correlation found for early neutrinos, which one could denote
 by $\rho_-$,
then probing the statistical significance of what one has found in terms of the product of these correlations.
Let us be satisfied here illustrating this strategy of analysis for the candidates listed in table 1.
First we address the issue of multiple candidate GRB partners for some of the neutrinos in table 1, by picking up the set
of GRB-neutrino candidates that maximizes the product of $\rho_+$ and $\rho_-$. This leads to selecting
as the 9 GRB-neutrino candidates  4  early neutrinos and 5 
late neutrinos\footnote{These 9 candidates selected with the criterion of maximizing the product $\rho_+ \cdot \rho_-$
differ from the 9 candidates with an asterisk in table 1 only in  one respect: they include  IC2/GRB100606A in place of IC2/GRB100605A.}. 
The resulting value of the product of correlations is $\rho_+ \cdot \rho_- = 0.812$
(with correlation of 0.981 for early neutrinos and of 0.828 for late neutrinos). We introduce a false alarm probability for this case
by producing as usual simulated data obtained by ramdomizing the times of observation of the 21 neutrinos in our sample (keeping as
usual their energies and directions unchanged), and seeing how frequently such simulations have at least 4 early-neutrino candidates
and at least 5 late-neutrino candidates, with the product of $\rho_+$ and $\rho_-$ greater or equal to $0.812$. We find that
this false alarm probability is 0.11$\%$.

\subsection{Early neutrinos with dual lensing}
Next we allow for dual lensing also for the analysis that focuses on early neutrinos.
For $\sigma_{d.l.} =5$degrees and for $\sigma_{d.l.} =10$degrees one does not pick up any additional GRB-neutrino candidates.
At $\sigma_{d.l.} =15$degrees one picks up a single additional early GRB-neutrino candidate, which is
IC30/GRB120709A. This is a rather intriguing dual-lensing candidate: as shown in Figure 4 it matches very naturally
the 5 early-GRB-neutrino candidates we started with (for $\sigma_{d.l.} =0$).

\begin{figure}[h!]

\includegraphics[scale=0.3]{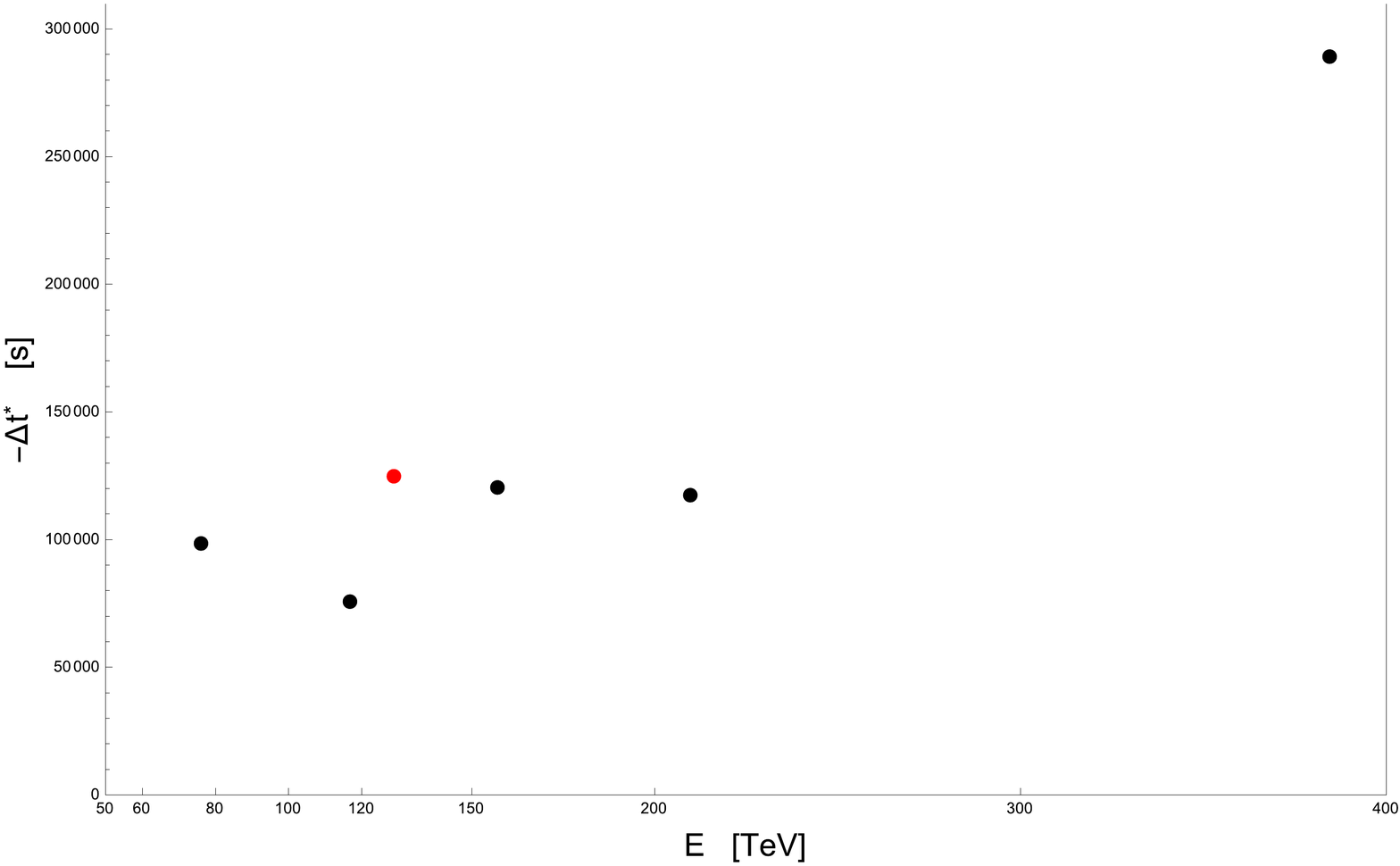}
\caption{Here in figure we just highlight the noteworthy consistency between the ``dual-lensing early neutrino candidate" (red) picked
 up at $\sigma_{d.l.}$ of 15 degrees and the 5 early-neutrino candidates (black) we already had without dual lensing ($\sigma_{d.l.} =0$).}
\end{figure}

Indeed adding this 6th candidate IC30/GRB120709A to the 5 early-GRB-neutrino candidates we started with
the correlation essentially remains unchanged: it changes from 0.945 to 0.937.
Computing our variation probability for this case, we find that only in 1.0$\%$ of cases by randomizing the times of detection
of the 12 neutrinos excluded at $\sigma_{d.l.} =0$
one would get accidentally a value of correlation $\geq 0.937$. Evidently this is of some (however tentative) encouragement for the
hypothesis of dual-lensing for early neutrinos. Note however that this is a case where on true data we picked up only one
dual-lensing candidate, and therefore the smallness of the variation of the correlation must be attribute not only to the
good match between the dual-lensing candidate and the other candidates, shown in Figure 4,
but also to the fact that with only 1 dual-lensing candidate it is difficult to produce large variations of a correlation
initially built out of 5 candidates. Simulations (obtained by time randomization for the 21 neutrinos in our sample)
producing several dual-lensing candidates would lead to a larger variation of the correlation even if each of these
dual-lensing candidates was a reasonably good match to the original 5 candidates.
Evidently this is a case where one should be particularly interested in checking the value of
what we labeled as the ``fixed-number variation probability", which we find to be of 39$\%$.
This value of 39$\%$ is still at least marginally encouraging for dual lensing of early neutrinos,
since it means that in 61$\%$ of cases simulations producing one and only one dual-lensing candidate lead to
values of correlation smaller than the value of 0.937 found in the analysis of the true data.

Increasing then $\sigma_{d.l.}$ to 20 degrees the numbers are even less intriguing.
For $\sigma_{d.l.} =20$degrees
one picks up two more candidates, which are IC51/GRB140414B and
IC22/GRB120118A, and one has a correlation of 0.826 for the total of 8 early-GRB-neutrino candidates
(the 5 candidates we already had
for $\sigma_{d.l.} =0$, the additional candidate picked up at $\sigma_{d.l.} =15$degrees, plus the 2 additional candidates
picked up at $\sigma_{d.l.} =20$degrees). The corresponding variation probability is  18$\%$,
while the fixed-number variation probability is  59$\%$.

\subsection{Late neutrinos with dual lensing}
Next we allow for dual lensing for the analysis that focuses on late neutrinos.
Something intriguing is immediately found by allowing
for $\sigma_{d.l.}$ of 5 degrees. One picks up two additional GRB-neutrino candidates,
which are IC39/GRB130707A
and IC46/GRB140129C.
Adding these 2 dual-lensing candidates  to the 7 late-neutrino candidates we started with (for $\sigma_{d.l.} =0$)
one has a total of 9 candidates for which the correlation between energy and $\Delta t^*$ is of 0.544, actually
higher than the correlation of 0.502 which we started with, for the 7
late-neutrino candidates already picked up at $\sigma_{d.l.} =0$. This is of course the type of  quantitative behavior
that supporters of dual lensing would want to see. In this particular case its significance is not very high, mainly as a result
of the fact that we started with a value of correlation, for $\sigma_{d.l.} =0$, which was not very high (0.502),
so it is not too difficult to pick up background neutrinos that accidentally look like dual-lensing candidates
producing an increase of the correlation. Indeed computing our variation probability for this case,
we find that in 19$\%$ of cases by randomizing the times of detection
of the 12 neutrinos excluded at $\sigma_{d.l.} =0$
one would get accidentally a value of correlation $\geq 0.544$.
As done in other analogous situations, we also consider more prudently the fixed-number variation probability,
which in this case comes out to be of  25$\%$.
So focusing on late neutrinos and $\sigma_{d.l.}$ of 5 degrees both in terms of our variation probability and in terms
of our fixed-number variation probability one finds that the true data are a bit more consistent with the dual-lensing hypothesis
than one would typically expect assuming instead that all neutrinos in our sample are background.

At $\sigma_{d.l.} =10$degrees one picks up a single additional late GRB-neutrino candidate, which is IC22/GRB120114B,
so the total number of candidates goes up to 10, and the corresponding 10-candidate correlation is of 0.404 (lower
than the 9-candidate correlation found at $\sigma_{d.l.} =5$degrees and actually also lower
than the 7-candidate correlation we started with at $\sigma_{d.l.} =0$).
Computing our variation probability for this case, we find that at $\sigma_{d.l.} =10$degrees in 39$\%$ of cases by randomizing the times of detection
of the 12 neutrinos excluded at $\sigma_{d.l.} =0$
one would get accidentally a value of correlation $\geq 0.404$.
The corresponding value of the fixed-number variation probability is of 46$\%$. Evidently the (however moderate) encouragement
for dual lensing for late neutrinos found at $\sigma_{d.l.} =5$degrees  vanishes already
at $\sigma_{d.l.} =10$degrees.

At $\sigma_{d.l.} =15$degrees one does not pick up any additional late GRB-neutrino candidate,
so one still has 10 candidates with correlation of 0.404. Recomputing the variation probability
for this case (with $\sigma_{d.l.} =15$degrees in the time-randomization analysis)
one finds for it the value of 37$\%$.
The corresponding value of the fixed-number variation probability is of 53$\%$.
So nothing much changes for the dual-lensing late-neutrino analysis
in going from $\sigma_{d.l.} =10$degrees to $\sigma_{d.l.} =15$degrees.

Finally increasing $\sigma_{d.l.}$ to 20 degrees one picks up 4 more candidates, for a total of 14,
the correlation goes down to 0.317, the variation probability is of 48$\%$, and the fixed-number
variation probability is also 48$\%$.

\section{Closing remarks}
This is the first ever truly quantitative phenomenological study centered on dual lensing,
which in itself should be viewed, in  our opinion, as an added value of our work.
Indeed previous studies of dual lensing had mostly focused on
the conceptual issues that still need to be addressed for
its full understanding \cite{freidelsmolin,ourlensing,transverseproceedings,stefanobiancolensing}.
We here essentially showed that these open conceptual issues do not obstruct the way for initiating
an associated phenomenological  program. Actually we would argue that progress in this phenomenological
effort needs most urgently not necessarily some theory work, but rather data of  improved quality. For example,
if the sample of high-energy neutrinos at our disposal had been larger we could have probed dual lensing
more finely, making room in  particular for the expected energy dependence of the ``dual-lensing angle".

Because of these limitations concerning the quality of data,
we expected this study to turn into a merely academic exercise, just introducing techniques of analysis
that might be valuable once indeed the quality of data improves. What we ended up finding goes somewhat
beyond the merely academic exercise, even though it is evidently just barely enough to provide motivation for
future related studies. One might want to take notice in particular of the fact
that our analysis of early neutrinos found for dual-lensing angle of 15 degrees
a result which is encouraging, though rather weakly. Similar remarks would apply to our findings
for late neutrinos at  dual-lensing angle of 5 degrees.
Moreover, it is intriguing that these results providing (however weak) encouragement for dual lensing
were exclusively found for small dual-lensing angles. For dual-lensing angle of 20 degrees all our results
provide no encouragement for dual lensing. Whether or not this difference between findings at smaller dual-lensing angles and
findings at larger dual-lensing angles is accidental (it is a quantitatively small difference anyway),
it serves us well in illustrating how in principle one could use an analysis such as ours not only to possibly establish
the presence of dual lensing, but also to estimate its magnitude.

We also believe that some of the observations and results we here reported will indeed be valuable also for studies
assuming that dual lensing is absent, as we anticipated in  our opening remarks. Estimating background for such studies is particularly challenging, and the strategy for estimating
background here proposed in subsection IV.A is a valuable step toward that goal. Moreover we here introduced 3 levels of analysis
which we feel should become a standard for similar studies, the level of the correlation for all neutrinos between energy
and $|\Delta t^*|$, the level of the correlation exclusively for late neutrinos between energy and $\Delta t^*$, and the level
of the correlation exclusively for early neutrinos between energy and $- \Delta t^*$.

\section*{Acknowledgements}

The work of GR  was supported  by funds provided by the National Science Center under the
agreement DEC- 2011/02/A/ST2/00294.
NL acknowledges support by the European Union Seventh Framework Programme (FP7 2007-2013) under grant agreement 291823 Marie Curie FP7-PEOPLE-2011-COFUND (The new International Fellowship Mobility Programme for Experienced Researchers in Croatia - NEWFELPRO), and also partial support from the H2020 Twinning project n$^o$692194, "RBI-TWINNING".

\end{document}